\documentclass[12pt]{article}
\usepackage{epsfig}
\usepackage{amsfonts}
\usepackage{amsopn}
\usepackage{amsmath}
\usepackage{verbatim,amsthm}

\title
{
\vskip -50 pt
\begin{flushright}
\normalsize\rm NORDITA-2009-30
\end{flushright}
\vskip 20 pt
$U(1)$-invariant membranes:
 the geometric formulation,
Abel and pendulum differential equations
}

\author{
 A. A. Zheltukhin $^{a,b,c}$\thanks{e-mail: aaz@physto.se} \
and M. Trzetrzelewski$^{d}$\thanks{e-mail: 33lewski@th.if.uj.edu.pl} \\ \\
$^a$ 1, Kharkov Institute of  Physics and Technology \\ Akademicheskaya St., Kharkov, 61108, Ukraine \\  \\
$^b$ Fysikum, AlbaNova, Stockholm University, \\
106 91, Stockholm, Sweden \\ \\
$^c$ NORDITA,  \\
Roslagstullsbacken 23, 106 91 Stockholm, Sweden \\ \\
 $^d$ M. Smoluchowski Institute of Physics, Jagiellonian University, \\
Reymonta 4, 30-059 Krak\'ow, Poland
}

\date{}

\begin{document}

\maketitle

\begin{abstract}
 The geometric approach to study the dynamics of $U(1)$-invariant membranes
is developed. The approach reveals an important role
of the Abel nonlinear differential equation of the first type with variable
coefficients depending on time and one of the membrane extendedness parameters.
The general solution of the Abel equation
is constructed. Exact solutions of the whole system of
membrane equations in the $D=5$ Minkowski space-time are found and classified.
It is shown that if the radial component of the membrane world vector
is only time dependent then the dynamics is described by the pendulum equation.

\end{abstract}

\section{U(1) invariant membranes}

The major role of membranes in M-theory poses the problem of their quantization
as one of the most fundamental in quantum field theory and strings \cite{Mem1,Mem2,Mem3,Mem4}.
In spite of the great progress obtained in studying the classical membrane dynamics
the question of its integrability \cite{integ1,integ2,integ3} is still open even for the case of
a free Dirac membrane in D-dimensional Minkowski
space-time  $x^{m}=(x^0,\vec{x})$ given by the Dirac action  \cite{Dirac}
\[
S=\int \sqrt{|G|}d\xi^3, \ \ \ G_{\alpha \beta}:
=\partial_{\alpha} x_{m}\partial_{\beta} x^{m},
\]
where $\xi^{\alpha}=(\tau,\sigma^r)$ (with $r=1,2$) and $G$ are the internal
coordinates and the determinant of the induced metric of the membrane world-volume respectively.
The principle obstacle is the three dimensional nonlinear nature of its
equations of motion
\begin{equation}\label{eom}
\dot{\mathcal{P}}^{m}=-\partial_r(
\sqrt{|G|}G^{r\alpha}\partial_{\alpha}x^{m}), \ \ \ \
\mathcal{P}^{m}=\sqrt{|G|}G^{\tau \beta}\partial_{\beta} x^{m}
\end{equation}
and the primary constraints
\begin{equation} \label{cons}
\mathcal{P}^{m} \partial_{r} x_{m} \approx 0, \ \ \
\mathcal{P}^{m}\mathcal{P}_{m}-\det G_{rs} \approx  0.
\end{equation}
A reduction of the 3-dim problem to an effective 2-dim one
could simplify the search for the general  solution of Eqs.
(\ref{eom}-\ref{cons}). In the paper \cite{JU1} Hoppe observed that
the presence of an additional $U(1)$ global symmetry, characterizing
the membrane shape, excludes the $\sigma^2$ parameter from the partial differential equations (PDE's)
(\ref{eom}) and (\ref{cons}). The general static solution of the
resulting two dimensional nonlinear problem was found in \cite{TZ} for D=2N+1
dimensional Minkowski space including  M-theory case D=11.

Here we propose a new approach to study the general time-dependent solutions of the
membrane equations of motion for D=5.
The $U(1)$ membrane is defined by the ansatz for the space
components $\vec{x}$ of its world vector
\begin{eqnarray}\label{ansatz}
\vec{x}^T=(m_1\cos\sigma^2,m_1\sin\sigma^2,m_2\cos\sigma^2,m_2\sin\sigma^2), \\
m_a=m_a(t,\sigma^1), \ \  \bold{m}:=(m_1,m_2).
\end{eqnarray}
 In the orthogonal  gauge
$\tau=x^0,  \ \ \   G_{\tau r}= -(\dot{\vec{x}} \cdot \partial_r\vec{x})=0$
the ansatz (\ref{ansatz}) results in the dynamics completely determined
by the following nonlinear differential equations
\begin{equation} \label{eq1}
\ddot{\bold{m}}=(\bold{m}^2\bold{m}^{'})^{'}-\bold{m}^{' 2}\bold{m}
\end{equation}
and  the constraints
\begin{equation}\label{eq2}
\dot{\bold{m}}^2+\bold{m}^2\bold{m}^{' 2}-1=0, \ \ \ \  \dot{\bold{m}}\bold{m}^{'}=0 ,
\end{equation}
where $\bold{m}^{'}$ and $\dot{\bold{m}}$ are partial derivatives of $\bold{m}$
with respect to $\sigma^1$ and $t$ respectively \cite{JU1} (see also \cite{TZ}).

It turns out that the static case $\dot{\bold{m}}=0$ is exactly
solvable and the general solution of (\ref{eq1}) and (\ref{eq2})
consists of two solutions \cite{TZ}. The first one is given by
\[
\bold{m}=\pm \sqrt{2\sigma^1}( \cos\psi_0,\sin\psi_0), \ \ \ \psi_0\in \mathbb{R},
\]
and describes a family of planes going through the origin of the Euclidean subspace
$\mathbb{R}^{4}$ of the Minkowski space.

The second solution is as follows
\[
\bold{m}(\sigma^1) = \bold{C}_1\sqrt{\frac{\sqrt{1+z^2}+1}{2}}+
\bold{C}_2\sqrt{\frac{\sqrt{1+z^2}-1}{2}},
\]
\[
\bold{C}_1=(C,D), \ \ \ \bold{C}_2=(D,-C),
\ \ \ z=\frac{2(\sigma^1 + \tilde c)}{C^2+D^2},
\]
where $C, D$ and  $\tilde c$ are the  integration constants,
and the vector function $\bold{m}$ yields the membrane surface which can be
visualized by rotating the hyperbola, lying in the
$x_1x_3$ plane, simultaneously in $x_1x_2$ and $x_3x_4$ planes.
Because of the scaling invariance $\tilde\sigma^1=b\sigma^1,\, \tilde m=\sqrt{b}m $
 of the static problem, we write out the general solutions for the
case $(\bold{C}_1\bold{C}_2)=0$ fixing the second integration
constant $\delta:=(\bold{C}_1\bold{C}_2)/c^2=0$ and simplifying the expression
of general solution with arbitrary  $\delta$.

There are also two time dependent, physically interesting solutions
\cite{TZ}. The first one corresponds to  a contracting, flat torus described by elliptic cosine
\begin{eqnarray}\label{tor1}
\bold{m}=\rho(t)\left(\cos n\sigma^1
,\sin n\sigma^1
 \right), \nonumber \\
\rho(t)=\frac{1}{\sqrt{n}}cn\left(\sqrt{2n}
t,\frac{1}{\sqrt{2}}\right), \ \ \ n \in \mathbb{Z}_{+}
\end{eqnarray}
and the second corresponds to a spinning, flat torus
\begin{eqnarray} \label{tor2}
\bold{m}=\rho(\sigma^1)\left(\cos\omega t ,\sin\omega t  \right), \nonumber    \\
\rho(\sigma^1)^2=\frac{1}{\omega^2}-(\pi-\sigma^1)^2\omega^2,
\end{eqnarray}
rotating with the frequency $\Omega \leq
\Omega_{max}= \frac{T^{1/3}}{\sqrt{\pi}}$,
where {T} is the membrane tension.

\section{The geometric approach and the Abel equation}

The case of the $U(1)$-invariant membranes (\ref{ansatz}), embedded in
the $D=5$ Minkowski space, is special since in that case  $\bold{m}^{'}$
and $\dot{\bold{m}}$ form a local basis attached to the effective
two-dimensional $\bold{m}$-surface (because of their linear independence provided
by the orthogonality constraint (\ref{eq2})).
An important consequence of the independence is that the constraints (\ref{eq2})
generate the dynamical equations (\ref{eq1}).
To see it one can, following \cite{JU1}, differentiate Eqs. (\ref{eq2})
and compare them with
Eqs. (\ref{eq1}) multiplied by $\dot{\bold{m}}$ and $\bold{m}^{'}$.

These observations  also imply that the dynamical equations (\ref{eq1}) preserve
the constraints (\ref{eq2}) which  may be treated
as the initial data constraints for Eqs. (\ref{eq1}) selecting a closed sector
of the non-static solutions. To analyze the sector we need to
study solutions of the constraints (\ref{eq2}).

To this end it is convenient to introduce a local moving frame in the
two dimensional $\bold{m}$-space
 formed by the unit orthogonal 2-vectors $\bold{n}_0$ and $\bold{n}_1$
\begin{equation}\label{reper}
\bold{n}_0=\frac{\dot{\bold{m}}}{\sqrt{1-\bold{m}^2\bold{m}^{'2}}},
\ \ \ \ \bold{n}_1=\frac{\bold{m}^{'}}{\sqrt{\bold{m}^{'2}}}, \ \ \
\ \bold{n}_i \cdot \bold{n}_j=\delta_{ij}.
\end{equation}
One can see that the constraints (\ref{eq2}) are equivalent to the orthonormality
conditions (\ref{reper}) for the local frame vectors
$\bold{n}_{i}(t,\sigma^1)$ (where $i,k=0,1$).

The introduction of the moving frame (or reper) opens a way to apply
 the famous Cartan approach \cite{Cartan}  to an alternative  
 formulation of our problem.
 The geometry of curved spaces \cite{Eise} in the Cartan approach
 is described not in terms of the original world coordinates $x^{m}$,
but in terms of the differential forms, associated with the vielbeins and connections
characterizing the spaces and defined by the Maurer-Cartan equations.
 In string theory the Cartan approach has been  firstly applied by
Regge and Lund \cite{RL1} (see also \cite{RL2}, \cite{RL3}) and was called the
geometric approach, or the embedding approach, because it describes the string worldsheet
as a two-dimensional surface embedded in the target space-time.

In the case at hand the above mentioned Cartan one-forms
$\omega_i= \omega_{i\mu}d\sigma^\mu$
and  $\mathcal{A}= \mathcal{A}_{\mu}d\sigma^\mu$, ($\sigma^\mu=t,\sigma^1$)
are defined by the following  equations
\begin{eqnarray}\label{difo}
\partial_{\mu}\bold{m}=\omega_{i\mu}\bold{n}_i, \ \ \
\partial_{\mu}\bold{n}_i=-\epsilon_{ik}\mathcal{A}_{\mu}\bold{n}_k,
\end{eqnarray}
 where $\epsilon_{ik}$ is the 2d antisymmetric Levi-Civita
tensor ($\epsilon_{01}= -\epsilon_{10}=1$), which encode the membrane evolution.
The connection  $\mathcal{A}_{\mu}$ is a  $U(1)$ gauge field belonging to the set of
gauge fields previously considered
 in the gauge reformulation of the Regge-Lund approach \cite{Z1,Z2,Z3,Z4,Z5,Z6}
\footnote{ The interpretation of the Maurer-Cartan equations in the non-linear
sigma-model theory
as constraints
 for the associated gauge fields was observed by Faddeev and Semenov-Tian-Shanskii \cite{FST}.}.

The integrability condition for the surface vector $\bold{m}$
\[
d^2 \bold{m}=0 \ \  \Rightarrow \ \
\partial_{[\mu}\omega_{i\nu]}+ \epsilon_{ik} \mathcal{A}_{[ \mu}\omega^k_{\nu ]}=0,
\]
gives the zero torsion conditions
\[
\partial_{\sigma}\sqrt{1-\bold{m}^2\bold{m}^{'2}}-\mathcal{A}_{\tau}\sqrt{\bold{m}^{'2}}=0,
\]
\begin{equation}
\partial_{\tau}\sqrt{\bold{m}^{'2}}+\mathcal{A}_{\sigma}\sqrt{1-\bold{m}^2\bold{m}^{'2}}=0
\label{mc}
\end{equation}
from which the connection components $\mathcal{A}_{\mu}$ can be read off explicitly.

The integrability condition  $d^2\bold{n}_i=0$  for $\bold{n}_i$
yields the  Maurer-Cartan equations for the connection $\mathcal{A}_{\mu}(t,\sigma^1)$
\[
F_{\mu\nu}=\partial_{[\mu}\mathcal{A}_{\nu]}=0 \Rightarrow \ \
\mathcal{A}_{\mu}=\partial_{\mu}\phi
\]
equivalent to the pure gauge condition for the abelian gauge field $ \mathcal{A}_{\mu}$
associated with the surface $\bold{m}(t,\sigma^1)$.
The scalar function  $\phi$
is defined by Eqs. (\ref{mc}) that may be presented in the form
\begin{equation}\label{mc1}
\frac{A^{'}}{B}=\dot{\phi}, \ \ \ \ \frac{\dot{B}}{A}=-\phi^{'},
\end{equation}
where
$A:=\sqrt{1-\bold{m}^2\bold{m}^{'2}}, \ \  B:=\sqrt{\bold{m}^{'2}}$.
Using (\ref{reper}) and (\ref{difo}) we find
the equations of motion (\ref{eq1}) to be equivalent to the following expansion
of $\bold{m}$
\begin{eqnarray}\label{mvec}
\bold{m}= \frac{\rho\rho^{'}}{B}\bold{n}_1 + \frac{\rho\dot\rho}{A}\bold{n}_0, \ \ \
\rho:= \sqrt{\bold{m}^{2}}=\sqrt{\vec{x}^2}
\end{eqnarray}
in the moving basis $\bold{n}_i$ (\ref{reper}).
 Thus, the components of $\bold{m}$
are expressed in terms of its length $\rho$, $A$ and $B$.
 It is easy to see that the projections
 of Eqs. (\ref{mvec}) on the linear independent 2-vectors  $\bold{m}^{'}$
and $\dot{\bold{m}}$ result in identities, so that  Eqs. (\ref{mvec}) turn out to be
 equivalent to the only one ellipse like relation
\begin{eqnarray}\label{roAB}
\left(\frac{\rho^{'}}{B}\right)^2 + \left(\frac{{\dot\rho}}{A}\right)^2 = 1
\end{eqnarray}
 which is the square of (\ref{mvec}).
The relation (\ref{roAB}) connects the functions  $\rho$,  $A$,  $B$ and
 is additional to the first of the constraints (\ref{eq2})
 \begin{equation}\label{constr2}
A^2 + \rho^2B^2=1.
\end{equation}
Then the problem is shifted to finding the solution of Eqs. (\ref{mc1}).

The first of Eqs. (\ref{mc1}) is presented in the simple form
\begin{eqnarray}\label{arcs}
\frac{(B\rho)^{'}}{ \sqrt{1-(B\rho)^2}}= -\frac{\dot{\phi}}{\rho}
\end{eqnarray}
 and its integration yields the following representations
\begin{eqnarray}
B= \frac{1}{\rho}\sin\lambda, \ \ \ A= \cos\lambda,  \label{BAsol}  \\
\lambda=- \int d\sigma^{1}\frac{\dot{\phi}}{\rho} + C_{0}(t)\label{lambdas},
\end{eqnarray}
for the functions $B$ and $A$.
The substitution of  the expressions for  $A$ and  $B$ (\ref{BAsol})
to the constraint (\ref{constr2}) yields an identity, so that  Eq. (\ref{constr2})
can be omitted. On the other hand the substitution of $A$ and $B$ to the second of
Eqs. (\ref{mc1}) and also in Eq. (\ref{roAB}) transforms them into
\begin{eqnarray}
\dot\lambda - \frac{\dot\rho}{\rho}\tan\lambda +\rho\phi^{'}=0, \label{lamb} \\
\left(\frac{\rho\rho^{'}}{\sin\lambda}\right)^2 +
\left(\frac{{\dot\rho}}{\cos\lambda}\right)^2 = 1. \label{rolamb}
\end{eqnarray}
As a result, the membrane dynamics in the geometric approach is
encoded  by the system of three coupled equations  (\ref{lambdas}),
(\ref{lamb}) and (\ref{rolamb}) for the three
functions $\lambda, \rho$ and $\phi$. In this system
the equation of motion (\ref{eq1})
is represented  by the differential equation (\ref{rolamb}) for the length
$\rho$ of the 2-vector $\bold{m}$.
The rest of the system, i.e. equations (\ref{lambdas}) and  (\ref{lamb}),
represent the constraints  (\ref{eq2})
accompanied by the equations of the moving reper.

There are a few ways to try to solve the nonlinear system
(\ref{lambdas}-\ref{rolamb}). At first one can observe that Eq. (\ref{rolamb})
does not include the function $\phi$ and is the exactly solvable biquadratic
(or quadratic if $\dot\rho=0$) equation for $\Lambda:=\tan\lambda$
\begin{eqnarray}\label{rolambalg}
{\dot\rho}^2\Lambda^4 - (1 -{\dot\rho}^2 -\rho^2\rho^{'2})\Lambda^2 + (\rho\rho^{'})^2 =0
\end{eqnarray}
  which presents  $\Lambda$ (and respectively $\lambda$) as the function of $\rho$
and its partial derivatives
\begin{eqnarray}\label{rolambroot}
\Lambda^2_{\pm}=[\, (1 -{\dot\rho}^2 -\rho^2\rho^{'2}) \pm
\sqrt{(1 -{\dot\rho}^2 -\rho^2\rho^{'2})^{2} -
4(\rho\dot\rho\rho^{'})^{2}}\, ]\frac{1}{{2\dot\rho}^2}\,\,.
\end{eqnarray}
The substitution of the solution (\ref{rolambroot}) in the remaining Eqs.
(\ref{lambdas}), (\ref{lamb}) transforms them into the second order differential
equation for the function $\phi$ with variable coefficients depending
on  $\rho$ and its derivatives.

An alternative way to derive the equation for $\phi$ is
to change Eq. (\ref{lambdas}) by the equivalent
differential equation and to unify  it together
with  Eq. (\ref{lamb}) in the form of system
\begin{eqnarray} \label{lambdaDer}
\lambda^{'}=-\frac{\dot{\phi}}{\rho},  \; \; \; \;
\dot\lambda =\frac{\dot\rho}{\rho}\tan\lambda -\rho\phi^{'},
\end{eqnarray}
 expressing the partial derivatives $\dot{\lambda}$ and $\lambda^{'}$
as the functions of $ \lambda, \rho, \dot\rho$ and the partial derivatives of $\phi$.
Then the integrability condition of the system  (\ref{lambdaDer}) yields the
2-dim {\it  linear} hyperbolic equation of the second order for $\phi$
\begin{eqnarray}\label{waveq}
\ddot\phi - \rho^2\phi^{''} -\rho^{-1}\dot\rho\dot\phi-\rho\rho^{'}\phi^{'} +
\rho(\rho^{-1}\dot\rho\Lambda)^{'} =0
\end{eqnarray}
 with variable coefficients and $\Lambda$ depending on $\rho, \rho', \dot\rho$,
as it follows from Eqs. (\ref{rolambroot}).
Of course, the explicit form of the $\Lambda$-roots  Eq. (\ref{rolambalg}) includes
 the square roots and yields a rather complicated dependence of the variable
coefficients of the wave-like equation (\ref{waveq}) on $\rho$ and its derivatives.
As a result, it complicates the solution of the wave equation (\ref{waveq}).
On the other hand a ${\it linear}$ character of Eq. (\ref{waveq}) opens a way to apply
 the general methods of the theory of PDE to seek the solution of this equation.

An alternative way to solve the geometric system  (\ref{lambdas}), (\ref{lamb}),
(\ref{rolamb}), encoding the membrane dynamics, is to present it
in terms of the above introduced function  $\Lambda:=\tan\lambda$ as follows
\begin{eqnarray}
\Lambda = \tan \left(- \int d\sigma^{1}\frac{\dot{\phi}}{\rho} + C_{0}(t) \right),
\label{Lambdas}
\\
\dot\Lambda=(1+ \Lambda^2)\left(\frac{\dot\rho}{\rho}\Lambda - \rho\phi^{'}\right),
\label{Abel} \\
{\dot\rho}^2 + \left(\frac{\rho\rho^{'}}{\Lambda}\right)^2 =\frac{1}{1+\Lambda^2},
\label{roABL}
\end{eqnarray}
where the representations $ A= \frac{1}{\sqrt{1+\Lambda^2 }}, \ \ B=
\frac{\Lambda}{\rho\sqrt{1+\Lambda^2 }} $ were used. Note that the
existing solutions (\ref{tor1}) or (\ref{tor2}) are consitent with
Eqs. (\ref{Lambdas} - \ref{roABL}).  To see this consider e.g.
solution (\ref{tor1}) which implies that $\rho'=0$ and $\phi^{'}=n$
so that Eq. (\ref{roABL}) gives $\Lambda=\pm
\frac{\sqrt{1-\dot{\rho}^2} \ }{\dot{\rho}}$. Choosing the $-$
branch and substituting $\Lambda$ to (\ref{Abel}) one finds a
differential equation consistent with (\ref{tor1}).
The  choice of the branch is important, i.e. if we take the $+$ branch
we will not find the consistency.

An interesting property of Eqs. (\ref{Lambdas} - \ref{roABL}) is the coincidence of
 Eq. (\ref{Abel}) with the well known
Abel equation having the cubic nonlinearity in $\Lambda$. Originally
the Abel equation of the first kind has its variable coefficients
depending only on the variable $t$. But, in our case the variable
coefficients of (\ref{Abel}) also depend  on the second variable
$\sigma^{1}$. Such dependence appears in the theory of extended
objects similar to strings/membranes and yields  information
about their dynamics. Thus, the derived
generalization of the Abel equation, arising from the geometric, approach opens
a new way for studying the nonlinear system of equations
(\ref{Lambdas}), (\ref{Abel}), (\ref{roABL}) and the $U(1)$
membrane physics.

 As a result, our statement is that the original description of the
membrane dynamics, given by the Eqs. (\ref{eq1}) and (\ref{eq2}), is reformulated
to the geometrical system of three equations (\ref{Lambdas}), (\ref{Abel}),
(\ref{roABL}) for functions $\Lambda, \rho$ and $\phi$,
 which  describe the effective 2-dimensional world sheet in terms of its
Cartan moving reper, metrics and connection fields.  The knowledge
of these functions enables to restore the effective membrane's
world vector $\bold{m}$ given by the expansion (\ref{mvec}). The
second statement is a partial encoding of the membrane dynamics by
the  {\it generalized} Abel equation (\ref{Abel}) with the $t$
{\it and $\sigma^{1}$} dependent coefficients. The appearance of
Abel equation gives new insights to the membrane dynamics
analysis. The third result of this geometric approach is the derivation
 of the {\it linear} wave-like equation (\ref{waveq}) associated with the $U(1)$ membrane.
The analysis based on this linear equation seems to be an alternative and promising.
However, in this  paper we focus our discussion on studying the
system (\ref{Lambdas}-\ref{roABL}) including the Abel equation (\ref{Abel}).

\section{Exact solutions and the pendulum equation}

Equations (\ref{Abel}) and (\ref{roABL}) may be easily solved for the
case of $\rho$ independent of time,  i.e. $\dot\rho=0$, because then
the Abel equation (\ref{Abel}) transforms into the Riccati equation that
has the general solution for $\lambda$ in the form
\begin{eqnarray}\label{Ricsol}
\lambda=- \rho\int dt\phi{'} + C_{1}(\sigma^{1}).
\end{eqnarray}
Then  Eq. (\ref{roABL}) shows that $\dot\lambda= 0$ which
implies  $\phi^{'}=0$, and $\dot{C_{0}}(t)=0$,
$\dot\phi =\omega=const$
because of (\ref{lambdas}). As a result, we obtain
\begin{eqnarray}\label{fisol}
\phi(t)=\omega t + b, \ \  \lambda=C_{1}(\sigma^1),\ \ \ (\omega,b,\tilde c\in \mathbb{R}),
\nonumber   \\
\rho^{2}(\sigma^{1})= \pm2\int d\sigma^{1}\sin C_{1}(\sigma^{1}) +\tilde c,
\ \  \rho C^{'}_{1}=-\omega.
\end{eqnarray}
The solution of Eqs. (\ref{fisol}) gives for $\rho(\sigma^{1})$ and $C_{1}(\sigma^{1})$
\begin{eqnarray}\label{roc1sol}
\rho(\sigma^1)^2=\frac{1}{\omega^2}-(\omega\sigma^1 + c)^2,\label{rocsol}
\\
C_{1}(\sigma^{1})=\pm \arcsin\sqrt{1- (\omega\rho)^2}.  \nonumber
\end{eqnarray}
 The solution (\ref{roc1sol}) coincides with the solution (\ref{tor2})
after using the periodicity condition.
It proves that (\ref{tor2}) is the general solution of the
toric membrane equations in the case $\frac{\partial \vec{x}^2}{\partial t}=0 $,
i.e. $\vec{x}^2=\bold{m}^{2}=\rho^2$  preserved in time.

The next solvable case of the Abel equation corresponds to $\rho$
independent of $\sigma^1$, i.e. $\rho^{'}=0$ that is equivalent to the
case $\frac{\partial \vec{x}^2}{\partial\sigma^1}=0$.
In this  case  Eq. (\ref{roABL}) shows that $\lambda^{'}= 0$, which
implies  $\dot\phi=0$ because of Eq. (\ref{lambdas}), and $\phi^{'}=const=\omega$
because of (\ref{Abel}), so that
\begin{eqnarray}\label{tor1fi}
\phi=\omega\sigma^1+a, \ \ \ \Lambda(t) =\tan C_{0}(t), \ \ (\omega,\, a \in \mathbb{R}).
\end{eqnarray}
As a result, Eqs. (\ref{Abel}-\ref{roABL}) take the form
\begin{eqnarray}
\rho\dot{C_{0}}=(\pm \sin C_{0}-\omega\rho^2),\label{AbeltorC} \\
\dot\rho=\pm \cos C_{0}. \ \ \ \ \ \  \ \   \label{Abeltorro}
\end{eqnarray}
 The differentiation of Eq. (\ref{AbeltorC}) results in the pendulum equation
for the function $C_{0}(t)$ coinciding with $\lambda$ (\ref{lambdas})
\begin{equation}\label{pendul}
\ddot{C_{0}}\pm 2\omega \cos C_{0}=0.
\end{equation}
Using Eq. (\ref{Abeltorro}) one can  present the pendulum equation
(\ref{pendul}) in the form of the total derivative
\begin{eqnarray}\label{conserv}
\frac{d}{dt}(\dot{C_{0}} + 2\omega\rho)=0 \Rightarrow
\dot{C_{0}} + 2\omega\rho=\tilde a.
\end{eqnarray}
Then the substitution of $\dot{C_{0}}$
 (\ref{AbeltorC}) to (\ref{conserv}) in combination with (\ref{Abeltorro})
gives the equation for the radial component $\rho(t)$ of the membrane world vector
\begin{equation}\label{elliptic}
{\dot\rho}^2 =1 -  {\rho}^2(\tilde a - \omega\rho)^{2}.
\end{equation}
The integration constant $\tilde a$ can be chosen to be
equal zero because for $\dot\rho(t_0)=1$, corresponding to the membrane motion
with the velocity of light ($\dot{\bold{m}}^{2}(t_0)=1$ at the moment $t_0$),
the radial component $\rho(t_0)$ vanishes,  as it follows from
Eqs. (\ref{AbeltorC}-\ref{Abeltorro}).
The general solution of Eq. (\ref{elliptic}) with $\tilde a=0$
is presented by  the elliptic integral
\begin{equation}\label{elliptint}
t=\int \frac{d\rho}
{\sqrt{1 - \omega^{2}\rho^{4}}} + const.,
\end{equation}
where the chosen sign plus in front of the integral correlates with
the opposite sign in the solution  $\Lambda= -\frac{\sqrt{1 - \dot\rho^{2}}}{\dot\rho}$.
The solution (\ref{elliptint}) coincides with the Jacobi elliptic
solution (\ref{tor1}), found in \cite{TZ},
and describes the toric membrane after taking into account
the periodicity condition for $\sigma^1$ implying  $\omega=n$.

We see that the geometric approach to the solution of
the nonlinear system (\ref{eq1}-\ref{eq2}), reformulating it into the system of
Eqs. (\ref{Lambdas}-\ref{roABL}), yields  the same particular time
dependent solutions as  Eqs. (\ref{eq1}-\ref{eq2}) \cite{TZ}.
The coincidence of the geometrical approach, presented by the
system (\ref{Lambdas}-\ref{roABL}),
shows its equivalence to the standard description of $U(1)$ membranes.

\section{On a general solution of \\
   the geometric approach equations}

Having the elliptic cosine solution or different particular solutions
for the case $\dot\rho\neq0$, could
we use them to construct the general solution of Eqs. (\ref{Lambdas}-\ref{roABL})
\footnote{The Referee is acknowledged
for sharpening this question.}?
We do not know the answer for this question for the time being.
However, there is some information connected with
 such a possibility which comes from the generalized Abel equation (\ref{Abel}).
 Studying this question in case of  the  Abel equation we find that
a  {\it sufficient} condition for a restoration
of its general solution in terms
of three known particular solutions is
 \begin{eqnarray}\label{4relat}
\frac{\rho^2\phi^{'}}{\dot\rho}=\chi(\sigma^1), \ \ \ \ \dot\rho\neq0,
\end{eqnarray}
where $\chi(\sigma^1)$ is an arbitrary function of the $\sigma^1$ membrane's parameter
\footnote{The case $\dot\rho=0$ reduces the Abel equation
to the Ricatti equation with the general solution (\ref{Ricsol})
considered in the previous section.}.

This observation  follows from interesting properties
of solutions of the original Abel equation of the first kind
\begin{equation}\label{Abel_t}
\dot{y}(t)= f_{3}(t)y^{3} +  f_{2}(t)y^{2} +  f_{1}(t)y  +  f_{0}(t)
\end{equation}
with variable coefficients $f_{\nu}(t), \,(\nu=0,1,2,3)$ depending on the argument $t$.
Because the r.h.s. of (\ref{Abel_t}) is a cubic polynomial, one can find its roots $y_{i}(t)$,
depending on $f_{\nu}(t)$, and present (\ref{Abel_t}) in the form
\begin{equation}\label{Abel_t1}
\dot{y}= f_{3}(y- y_{1})(y- y_{2})(y - y_{3} ), \ \ \  f_{3}(t)\neq0.
\end{equation}
It is clear that  the roots $y_{i}(t)$ are particular solutions of the Abel equation,
 if its variable coefficients $ f_{\nu}(t)$ are connected by the conditions
\begin{equation}\label{Ab_tcond}
\dot{y}_{i}(t)=0
\end{equation}
which means that $ y_{i}=const$.
For such
coefficients $ f_{\nu}(t)$, the general solution of the Abel
equation (\ref{Abel_t}) is given by the expression
\cite{SAAT}\footnote{Take into account misprints in the
representation of the general solution in \cite{SAAT}.}
\begin{equation}\label{Ab_tgensol}
\prod^{3}_{i=1}
(y- y_{i})^{\alpha_{i}}= \kappa e^{\int dt f_{3}(t)},
\end{equation}
where $\kappa$ is an arbitrary integration constant and the constants $\alpha_{i}$
are defined by the roots $y_{i}$. We find the constants solving the set of algebraic equations
\begin{eqnarray}
\alpha_{1}+\alpha_{2}+\alpha_{3}=0, \ \ \
\alpha_{1}y_{2}y_{3}+ \alpha_{2}y_{3}y_{1} +\alpha_{3}y_{1}y_{2}=1,
\nonumber
 \\
\alpha_{1}(y_{2}+ y_{3})+ \alpha_{2}(y_{3}+ y_{1})+\alpha_{3}(y_{1}+ y_{2})=0.
\label{Ab_tgenalg}
\end{eqnarray}
One can try to use the representation (\ref{Ab_tgensol})
 to construct the general solution of (\ref{Abel}) despite the fact that its
coefficients depend not only on $t$ but also on the membrane
 {\it extendedness}   coordinate $\sigma^1$. A necessary condition to
preserve the representation (\ref{Ab_tgensol}) is to treat the
constants $\alpha_{i}$ as the functions $\alpha_{i}(\sigma^1)$
satisfying the same conditions (\ref{Ab_tgenalg}).
Moreover, Eqs. (\ref{Ab_tgenalg}) for the
coefficients in Eq. (\ref{Abel})
have to be consistent with the remaining equations (\ref{Lambdas}) and (\ref{roABL}).

We observe that the two roots of the cubic polynomial in the r.h.s.
of the Abel equation (\ref{Abel})
\begin{equation}\label{polynom}
{\cal P}(\Lambda)= (1+ \Lambda^2)\left(\frac{\dot\rho}{\rho}\Lambda - \rho\phi^{'}\right)
\end{equation}
 are the complex conjugate numbers
\begin{equation}\label{rootpm}
\Lambda_{\pm}=\pm i
\end{equation}
which actually are particular solutions of (\ref{Abel}), and they do not create  any
additional conditions on the functions $\rho$ and $\phi$.
However, the third root $\Lambda_{3}$ of the polynomial ${\cal P}(\Lambda)$ (\ref{polynom})
\begin{equation}\label{root3}
\Lambda_{3}= \frac{\rho^2\phi^{'}}{\dot\rho},  \ \ \  \dot\rho\neq0
\end{equation}\label{root3_cond}
yields the new additional condition $\dot\Lambda_{3}=0$,
connecting  $\rho$ and $\phi$
\begin{equation}\label{root3cond1}
\dot\Lambda_{3}=\frac{\partial}{\partial t} (\frac{\rho^2\phi^{'}}{\dot\rho})=0 \ \ \                           \rightarrow \ \ \ \frac{\rho^2\phi^{'}}{\dot\rho}=\chi(\sigma^1),
\end{equation}
where the  integration "constant" $\chi(\sigma^1)$ is the
previously mentioned function in (\ref{4relat}).
Because the fulfilment of the condition (\ref{root3cond1}) is the  {\it necessary}
condition for the presentation of the general solution in the form (\ref{Ab_tgensol}),
this condition  has to be considered as a new constraint for $\rho$ and $\phi$
additional to Eqs. (\ref{Lambdas},\, \ref{roABL}).

Assuming that Eq. (\ref{root3cond1}) is actually  consistent with the
remaining equations  (\ref{Lambdas},\, \ref{roABL}), one can solve the algebraic
system (\ref{Ab_tgenalg}) and obtain
\begin{eqnarray}
\alpha_{+}= \frac{i}{2(\Lambda_{3}-i)}, \ \ \
\alpha_{-}=(\alpha_{+})^{*}= \frac{-i}{2(\Lambda_{3}+i)}, \nonumber  \\
\alpha_{3}= -(\alpha_{+}+\alpha_{-})= \frac{1}{1 + \Lambda_{3}^{2}} . \label{roots}
\end{eqnarray}
The substitution of the solution (\ref{roots}) in the representation (\ref{Ab_tgensol})
yields the desired expressions for the general solution $\Lambda(t,\sigma^1)$
of the membrane Abel equation  (\ref{Abel}) in the complex form
\begin{eqnarray}\label{funcsol}
\frac{\Lambda-\Lambda_{3}}{\sqrt{1 + \Lambda^{2}}}
\left ( \frac{\Lambda+i}{\Lambda-i} \right )^{\frac{-i\Lambda_{3}}{2}}
= (\kappa\rho)^{(1 + \Lambda_{3}^{2})}
\end{eqnarray}
where $\kappa$ is now an arbitrary function of $\sigma^1$.
Taking into account that the general physical solution implies the reality of $\Lambda$ in
(\ref{funcsol}) and using the relation
\begin{eqnarray}\label{moduli}
 \frac{\Lambda+i}{\Lambda-i}= e^{2i \arctan(\Lambda^{-1})}
\end{eqnarray}
one can rewrite the expression (\ref{funcsol}) in the real form
\begin{eqnarray}\label{realsol}
\frac{\Lambda-\Lambda_{3}}{\sqrt{1 + \Lambda^{2}}}
= (\kappa\rho)^{(1 + \Lambda_{3}^{2})} e^{-\Lambda_{3}\arctan(\Lambda^{-1})},
\end{eqnarray}
where  $\Lambda_{3}= \frac{\rho^2\phi^{'}}{\dot\rho}$ in accordance with
 (\ref{root3}).
So, we obtain the desired general solution of Eq. (\ref{Abel}), accompanying
the remaining Eqs. (\ref{Lambdas},\ \ref{roABL}),  in the form
of the non-algebraic functional relation connecting the functions
$\Lambda, \rho$ and $ \phi$ .

It proves that the general solution of the Abel equation (\ref{Abel}) is expressed
in terms of the three particular solutions, if the constraint  (\ref{4relat}) is satisfied.

An example of an explicit solution, encoded by the representation (\ref{realsol})
and satisfying the constraint  (\ref{4relat}), is given by the choice
\begin{eqnarray}\label{part_genercond}
\Lambda_{3}=0 \ \ \    \rightarrow  \ \ \phi^{'} =0, \ \ \chi=0
\end{eqnarray}
which reduces the implicit equation (\ref{realsol}) to a simple explicit form
\begin{eqnarray}\label{prt_genersol}
\Lambda=\pm\frac{\kappa\rho}{\sqrt{1-(\kappa\rho)^2}}.
\end{eqnarray}

Unfortunately, the condition (\ref{4relat}) is very bounding and excludes
the elliptic solution (\ref{elliptint}). It follows from the constraint (\ref{4relat})
matching with the conditions
\begin{eqnarray}\label{elipcond}
\rho^{'} =\Lambda^{'}=\dot\phi=0,
\end{eqnarray}
fixing the elliptic solution, that yields either
solution (\ref{part_genercond}) or the solution
\begin{eqnarray}\label{eliprestr}
\frac{d(\rho^{-1})}{dt}=-\frac{\phi^{'}}{\chi}=const.,
\end{eqnarray}
 both of which are different from the elliptic cosine solution.

Thus, the problem appears in the realization of the formulated way to construct
the general solution of the geometric system (\ref{Lambdas}-\ref{roABL}).
 However, in this way we observe another interesting property
 of the Abel equation (\ref{Abel}) which might be characterized as its
duality with the Ricatti equation.
This property is discussed in the next section.

\section{A duality between the Abel and \\
 Ricatti membrane equations}

The observation is that  Eq. (\ref{Abel}), considered as the Abel equation for $\Lambda$
with the coefficients depending on $\rho$ and $\phi$,
transforms to the
Ricatti equation for the function $\rho$ with the coefficients
depending on $\Lambda$ and $\phi^{'}$
\begin{eqnarray}\label{Ric_Ab}
\dot\rho= \rho\Lambda^{-1}
\left(\frac{\dot\Lambda}{\sqrt{1+\Lambda^{2}}} +\rho\phi^{'}\right).
 \end{eqnarray}
After the substitution  $r=1/\rho$  the Ricatti equation (\ref{Ric_Ab}) transforms into
the linear differential equation  (cp. (\ref{lambdaDer}))
\begin{eqnarray}\label{linRic_Ab}
\frac{\partial(rsin\lambda)}{\partial t}=- \phi^{'}cos\lambda
\end{eqnarray}
and its general solution is obtained by the quadrature
\begin{eqnarray}\label{solRic_Ab}
r =\frac{1}{sin\lambda} \left( {\bar C}(\sigma^1) -\int dt\phi^{'}cos\lambda \right ),
\end{eqnarray}
where ${\bar C}(\sigma^1)$ is the integration constant.
Thus, the general solution $\Lambda$
of the Abel equation studied above is encoded in the integral relation
\begin{eqnarray}\label{Abel_space}
\frac{\Lambda}{\rho\sqrt{1+\Lambda^{2}}}=
-\int dt \frac{\phi^{'}}{\sqrt{1+\Lambda^{2}}}+ {\bar C}(\sigma^1)
\end{eqnarray}
instead of the  discussed non-algebraic functional equation (\ref{realsol}).
The extension captures the whole space of the Abel equation solutions.
One can  see, e.g. that the particular solution (\ref{prt_genersol}) of Eq. (\ref{realsol})
is extracted from the general solution (\ref{Abel_space}) by the
condition $\phi^{'}=0$ and the identification ${\bar C}(\sigma^1)=\kappa(\sigma^1)$.
The desired  elliptic solution (\ref{elliptint}) is respectively extracted
by the conditions $\phi^{'}=\omega, \ \  \Lambda=-\sqrt{1- \dot\rho^{2}}/\dot\rho, \ \
\sqrt{1+\Lambda^{2}}=1/\dot\rho$,
whose substitution into (\ref{Abel_space}) transforms it into the equation  (\ref{elliptic})
\begin{eqnarray}\label{Abel_ellip}
{\dot\rho}^2 =1 -  {\rho}^2({\bar C} - \omega\rho)^{2}
\end{eqnarray}
 which results in  the elliptic cosine  solution (\ref{elliptint})
after identification ${\bar C}= \tilde a =0$.

Taking into account that the integral relation (\ref{Abel_space}) may be considered 
as the general solution for the membrane length $\rho$, expressed in terms 
of $\lambda$ and $\phi^{'}$,
one can present the geometric system (\ref{Lambdas}-\ref{roABL}) in 
the $\lambda$-representation  $(\lambda=\arctan\Lambda)$ as
\begin{eqnarray}
(cos\lambda)^{'} =
\dot\phi\left( {\bar C}(\sigma^1) -\int dt \phi^{'} cos\lambda\right),  \label{ugol}
\\
\frac{1}{\rho}=\frac{1}{sin\lambda} \left( {\bar C}(\sigma^1) -
\int dt\phi^{'}cos\lambda \right ), \label{dlina}
\\
\left(\frac{\rho\rho^{'}}{\sin\lambda}\right)^2 +
\left(\frac{{\dot\rho}}{\cos\lambda}\right)^2 = 1.  \label{svyaz}
\end{eqnarray}

A new  property of the representation is that Eq. (\ref{ugol}) 
takes  the form of the integro-differential equation for $cos\lambda$,
 which defines $\lambda$ as a function of only the partial derivatives 
of $\phi$ and the integration constant ${\bar C}(\sigma^1)$.
The substitution of the supposed $\lambda$-solution into Eq. (\ref{dlina}) 
will also express  the length $\rho$  as a function of  only the derivatives of $\phi$ 
and $\bar C$. The substitution of these $\lambda$ and $\rho$ 
to the last Eq. (\ref{svyaz}) will yield  the equation for the last unknown
 function of the geometrical approach, the phase $\phi$ (cp. Eq. (\ref{waveq})).              

The system (\ref{ugol}-\ref{svyaz}) is a final point in  our discussion 
of the geometric formulation in this paper. 
It leaves the problem of the general solution of the $U(1)$ membrane equations
still open.  Nevertheless, the system  (\ref{ugol}-\ref{svyaz}) seems a promising
starting point for new attempts to solve the problem.

\section{Conclusion}

 The investigation of classical and quantum dynamics of the non-linear
sigma-models created a powerful group theory methods for their study
\cite{FST, FaTa}. The Regge-Lund geometric  approach  \cite{RL1,RL2,RL3}, based on
the classical theory of embedded surfaces  \cite{Eise}, and its
reformulation in terms of gauge
fields \cite{Z1,Z2,Z3,Z4,Z5,Z6}
were applied here in search of the general
 solution of  the $U(1)$ membrane equations in the five dimensional
Minkowski space. We revealed that the membrane nonlinearities are
partially encoded by the cubic polynomial of the known Abel
differential equation of the first type,
but with the variable coefficients depending on the membrane's
extendedness parameter $\sigma^1$ in addition to the original time
dependence.
We found that  the cases
when the radial component  $\rho$ of the membrane world vector depends
only on the time $t$ or only on the  extendedness parameter $\sigma^1$,
the membrane equations are exactly solvable.
The obtained solutions are proven to be  the general ones for such types
of the $\rho$ dependencies, and they coincide with the previously found
solutions \cite{TZ}.
 For the case of  $\rho$ dependent only on time the dynamics of membranes
turns out to be connected  with the well known nonlinear differential pendulum
equation.

In search of an alternative approach to the solution of the $U(1)$ membrane equations
we also derived  a  2-dim {\it  linear} hyperbolic
equation of the second order (for a phase $\phi$)
with variable coefficients, depending on the membrane length $\rho$ and its
partial derivatives  $\rho', \dot\rho$.
 The linear character of this equation allows us to apply the general
methods of the PDE theory to seek the desired general solution.

Some attempts were undertaken here to find the general solution of the whole
system of nonlinear equations in the considered geometric approach.
The {\it sufficient} criteria to restore the general solution of the membrane
 Abel equation in terms of its three particular solutions was formulated.
Using this criteria we present the general solution of the membrane
Abel equation in the form of a non-algebraic functional equation,
connecting the geometric approach functions. However, this criteria
forbids the elliptic solution of the Abel equation. Then we observed
that the Abel equation, interpreted as an equation for the membrane
length, coincides with the Ricatti equation and we found its general
solution by quadrature. The general solution of the  Ricatti
equation encodes the whole space of the Abel equation solutions. It
seems that the derivation of the integral representation for the
membrane length partially simplifies the  whole system of membrane
equations and gives new tools to solve the problem.
Thus,  the presented geometrical reformulation of the $U(1)$ membrane
mechanics gives a new promising information which will help in the search for its solvability.

\section*{Acknowledgements}

We are grateful to J. Hoppe for valuable collaboration and
support.  AZ would like to thank  J.
{\AA}man, I. Bengtsson, R. Myers, N. Pidokrajt and N. Turok for useful discussions,
 and Fysikum at
Stockholm University, Department of Mathematics of Royal Inst. of
Technology KTH Stockholm, Nordic Inst. for Theoretical Physics
Nordita and Perimeter Inst. for Theoretical Physics for kind hospitality.
 MT would like to thank A. Wereszczy\'nski for discussions and the Albert
Einstein Institute for hospitality. This research was supported in part by  Nordita,
Knut and Alice Wallenberg Foundation, Marie Curie Research Training Network
ENIGMA (contract MRNT-CT-2004-5652)  and  Perimeter Institute for Theoretical Physics.

\end{document}